\documentclass[a4paper,11pt]{article}
\usepackage{amsmath,amsfonts,latexsym}
\usepackage{amsthm}
\usepackage{amssymb}
\usepackage[dvips]{graphicx}
\usepackage{array}

\makeatletter

\newcommand{\C}{\mathbb C}
\newcommand{\R}{\mathbb R}
\newcommand{\Z}{\mathbb Z}
\newcommand{\N}{\mathbb N}

\theoremstyle{plain}
\newtheorem{Thm}{Theorem}

\newtheorem{Pro}[Thm]{Proposition}
\newtheorem{Cor}{Corollary}[Thm]
\newtheorem{Lem}[Thm]{Lemma}

\theoremstyle{remark}

\newtheorem{Rem}[Thm]{\sc Remark}

\newcommand{\ps@taiplain}{%
     \renewcommand{\@oddhead}{\centerline{\sf {\small Remarks on $\mathcal{PT}$-pseudo-norm in
     $\mathcal{PT}$-symmetric quantum mechanics}}}%
     \renewcommand{\@evenhead}{\@oddhead}%
     \renewcommand{\@oddfoot}{\centerline{\thepage}}
     \renewcommand{\@evenfoot}{\@oddfoot}}
\makeatother     

\begin{document}

\thispagestyle{plain}

\title{\bf Remarks on $\mathcal{PT}$-pseudo-norm in $\mathcal{PT}$-symmetric quantum mechanics}
\vspace{-2cm}
\author{
{\em TRINH Duc Tai}
\thanks{Department of Mathematics, Teacher Training College of Dalat,
  29 Yersin, Dalat, Vietnam / tel.
84 63 827344 / fax. 84 63 834732/  {\sc email:}
tductai@hcm.vnn.vn}'\thanks{The Abdus Salam International Centre
for Theoretical Physics, Strada Costiera 11, Trieste 34014, Italy/
{\sc email:} tductai@ictp.trieste.it}} \vspace{-2cm}
\date{\today}

\maketitle

\pagestyle{taiplain}

\begin{abstract}
This paper presents an underlying analytical relationship between
the $\mathcal{PT}$-pseudo-norm associated with the $\mathcal{PT}$-symmetric
Hamiltonian $H = p^2+ V(q)$ and the Stokes multiplier of the
differential equation corresponding to this Hamiltonian. We show
that the sign alternation of $\mathcal{PT}$-pseudo-norm, which has been
observed as a generic feature of the $\mathcal{PT}$-inner product, is essentially
controlled by the derivative of Stokes multiplier with respect to
eigenparameter.
\end{abstract}



\section{Introduction}

Since the late 1990's, the notion of $\mathcal{PT}$-symmetric Hamiltonians in
quantum physics is addressed in many works after its first
appearance in a publication of Bender and Boettcher
\cite{BenBoe98} concerning a conjecture of Bessis and
Zinn-Justin.

By dropping the requirement of Hermiticity but of course keeping the invariance by the
Lorentz group, one opens on a significantly larger class of
Hamiltonians satisfying a weaker hypothesis, namely
the $\mathcal{PT}$-invariance. This more flexible condition amounts to
the commutability of the Hamiltonian $H$ and the composite operator
$\mathcal{PT}$ whose components consist of one linear operator
$\mathcal{P}$ (for \emph{parity}) and another anti-linear
$\mathcal{T}$ (for \emph{time-reversal}).

It has been showed that, despite the lack of Hermiticity, many
$\mathcal{PT}$-symmetric Hamiltonians still have a whole real and
bounded from below discrete spectrum (see
\cite{BenBerry002,BenBoe98-2,CGM80,DP98-1,DT00,Dorey,LevaiZnojil,Shin002}).
However, a controversy has merged among researchers due to the
fact that the space of states may be no longer a Hilbert space, at
variance with  traditional quantum mechanics.

Recently, many studies have been devoted to the establishment of a
conventional mathematical structure for $\mathcal{PT}$-symmetry
quantum mechanics (see \cite{Ben0,Ben1,Mostaf2,Mostaf3}). One of
the most important considerations is to equip the space of states
associated with a given Hamiltonian with a certain inner product,
so suitably that the norm induced by this inner product is
positive definite.

For Hermitian Hamiltonians on the real axis $\R$, such a space is
normally known as the Hilbert space $L^2(\R)$, whose inner product
is nothing but the usual one. The norm induced from this actually
positive definite inner product is interpreted as a probability
density in the space of states.

Such an ideal model seems to be no longer available whenever the
Hamiltonians are non-Hermitian. However, the specialists in
the field
have recently  succeeded in constructing a consistent physical
theory for \emph{unbroken} $\mathcal{PT}$-symmetric Hamiltonians.
The most essential aspect in their investigations is the
appearance of a linear  \emph{``charge"}
 operator $\mathcal{C}$  whose action enables the
so-called $\mathcal{PT}$-pseudo-norm
to switch its sign in a suitable way, so that the induced
$\mathcal{CPT}$-norm is now positive definite.

This operator $\mathcal{C}$ is nowdays a central subject of
interest. Recently, in an attempt to formulate
$\mathcal{PT}$-symmetric quantum theories, Bender \emph{et al}
have obtained many significant results through explicit
calculation of the operator $\mathcal{C}$ for some special
$\mathcal{PT}$-symmetric Hamiltonians (see
\cite{Ben0,Ben1,BenBerry003}).  As a rule, the  construction
relies on the observation  of a so-called quasi-parity quantum
number \cite{Bag01,LevaiCan,Znoj0104}, so
 that the  $\mathcal{PT}$-pseudo-norm exhibits a generic
sign alternation, upon the eigenstates of Hamiltonians.

This paper, which is motivated by these observations, aims mainly to
justify the indefiniteness of the $\mathcal{PT}$-inner product for a class of
unbroken $\mathcal{PT}$-symmetric Hamiltonians. By using some
classical results on Stokes multipliers (see \cite{Sibuya}, ch.
$5,6$), whose zeros are exactly the eigenvalues of the considered
Hamiltonians,  we show an analytical relation between
the $\mathcal{PT}$-pseudo-norm and the derivative with respect to
the eigenparameter of one of these Stokes multipliers, from which the sign
alternation occurs as a natural consequence.

This paper is organized as follows. In  section 2 we recall
some simple notions and facts on the Sturm-Liouville eigenvalue
problem associated with a class of complex Hamiltonians. A strong
connection between our problem and Sibuya's theory on linear
differential equations is formulated for our next purposes.
Section 3 contains our main results which allow to establish a
relationship between the $\mathcal{PT}$-pseudo-norm and the derivative in the
eigenparameter of a convenient Stokes multiplier. This
 gives a clear explanation for the indefiniteness of the
 $\mathcal{PT}$-pseudo-norm,  thus justifying what has been observed in various
 works (e.g.,  \cite{Ben0}).
Finally, in the conclusion we suggest some further considerations
and discuss a degenerate case, where the $\mathcal{PT}$-symmetry
may be spontaneously broken by the vanishing of the
$\mathcal{PT}$-pseudo-norm at some degenerate states.

\section{Sturm-Liouville eigenvalue problem associated to a Hamiltonian}

We consider a non-Hermitian Hamiltonian
\begin{equation}\label{hamil}
    H = p^2 + V(q)
\end{equation}
where $p$ stands for the operator $-i\frac{d}{dq}$ and
\begin{equation}\label{dathuc}
   V(q) := -[(iq)^m + a_1(iq)^{m-1} + \cdots + a_{m-1}(iq) + a_m]
\end{equation}
is a polynomial of degree $m$ in the variable $iq$ with
\emph{real} coefficients $a_j \in \R$.

These assumptions induce that the complex-valued
potential function $V(q)$ satisfies the following relation,
\begin{equation}\label{ptdx}
    \overline{V(-\overline{q})} = V(q)
\end{equation}
(where $\overline{q}$ stands for the complex conjugate of $q$),
which is broadly known as the $\mathcal{PT}$-symmetry, or more
geometrically as the invariance property of $V(q)$ under the
\emph{real} conjugation.

Recall that, as usual, the combination $\mathcal{PT}$  stands for
the composite operator of $\mathcal{P}$ (\emph{parity} operator)
and $\mathcal{T}$ (\emph{time-reversal} operator) whose actions on
the $(p,q)$-space are generally determined as follows:
$$\mathcal{P}: \left\{
\begin{array}{l}
q \mapsto - q\\
p \mapsto -p
\end{array}
\right. \quad \rm {and}\quad \mathcal{T}: \left\{
\begin{array}{l}
q \mapsto  \overline{q}\\
p \mapsto -\overline{p}
\end{array}
\right.
$$

By definition, $\mathcal{PT} (=\mathcal{TP})$ is not a linear but an anti-linear
operator. Its action on a wave function $\phi(q)$ is simply
given by
\begin{equation}\label{dncuapt}
    \mathcal{PT}\phi(q) = \overline{\phi(-\overline{q})}.
\end{equation}
Consequently, a function $\phi(q)$ is called
$\mathcal{PT}$-\emph{symmetric} if it remains unchanged under the
action of the operator $\mathcal{PT}$ in (\ref{dncuapt}). It is
obvious that the Hamiltonian $H$ in (\ref{hamil}) is
$\mathcal{PT}$-symmetric, i.e.
\begin{equation}\label{hamilvapt}
    \mathcal{PT}H = H\mathcal{PT}
\end{equation}
provided that $V(q)$ is $\mathcal{PT}$-symmetric. And this is the case when all
coefficients $a_j$ in (\ref{dathuc}) are real.

\subsection*{$\mathcal{PT}$-inner product and
$\mathcal{PT}$-pseudo-norm.}

In the sequel, we shall determine a space of functions on which the action
of our given Hamiltonian is meaningful. We note that the usual set of
functions considered for most (Hermitian) Hamiltonians on the real
line is $L^2(\R)$, in relation to its useful  Hilbert
space structure. In the context of non-Hermitian Hamiltonians, defining such a
space with an appropriate algebraic structure
has not been yet completely settled in general.

Consider the Hamiltonian $H$ in (\ref{hamil}) with the assumption
that all $a_j$ in (\ref{dathuc}) are real, so that $H$ is
$\mathcal{PT}$-symmetric. For a fixed $m\in \N$, $m \geq 2$, we
denote by
 $\mathfrak{H}$
the complex vector space  of all entire functions $f(q)$ which are exponentially
vanishing at infinity in both of the two  open sectors
\begin{equation}\label{sectortraiphai}
    \begin{array}{lll}
       & & S_l := \left\lbrace \big\vert \arg(X) +
       \frac{\pi}{2}+\frac{2\pi}{m+2} \big\vert < \frac{\pi}{m+2}
       \right\rbrace \\ \vspace{3mm}
       {\rm and} & & S_r := \left\lbrace \big\vert \arg(X) +
       \frac{\pi}{2}-\frac{2\pi}{m+2} \big\vert < \frac{\pi}{m+2}
       \right\rbrace \\
   \end{array}
\end{equation}
This space  $\mathfrak{H}$   can be considered as a vector
subspace of the space of square integrable functions in both the
two  neighborhoods of infinity ${S_l}$ and ${S_r}$. By virtue of
the involution of $\mathcal{PT}$ and the definition of
$\mathfrak{H}$, we have immediately
\begin{equation}\label{tamtam}
\mathcal{PT}(\mathfrak{H})= \mathfrak{H}.
\end{equation}

The following statement, which can be checked directly, asserts
the symmetry of the discrete spectrum of a
$\mathcal{PT}$-symmetric Hamiltonian.
\begin{Pro}\label{doixungxung}
    Let $\mathfrak{G}$ be a certain vector space of functions
    satisfying $\mathcal{PT}(\mathfrak{G}) \subset \mathfrak{G}$.
    Then the set of eigenvalues of a $\mathcal{PT}$-symmetry
    Hamiltonian $H$ acting on $\mathfrak{G}$ is invariant under
    complex conjugation.
\end{Pro}

We are going to introduce a space in which each eigenfunction
of the non-Hermitian Hamiltonian $H$ is associated with a real
number, usually (but a bit abusively) called its $\mathcal{PT}$-pseudo-norm.

Instead of considering the real axis, as it is the case for the usual
 norm  in $L^2(\R)$, a slightly different curve will be involved in
 order to be
consistent with our goals. Consider an endless curve $\gamma$
starting from infinity in the sector $S_l$ and ending also at
infinity but in the different sector $S_r$. Note that for a
given function $h(q)\in \mathfrak{H}$, which is thus  holomorphic and integrable at
infinity in  ${S_l}\cup {S_r}$, the value of the integral
$\displaystyle \int_{\gamma}h(q)dq$ remains unchanged when $\gamma$ is
deformed continuously  such that both of its endpoints still lie in
$S_l$ and $S_r$ respectively.

For the sake of simplicity, we shall take $\gamma$ to be symmetric
so that, up to its orientation, the mapping $q \mapsto
-\overline{q}$ of the complex plane $\C$, which is nothing but the
mirror symmetry with respect to the imaginary axis, leaves
$\gamma$ invariant. Precisely, it will be convenient to define
$\gamma$ by:
\begin{equation}\label{chongamma}
    \gamma := \gamma_r - \gamma_l ,
\end{equation}
where $\gamma_r$ and $\gamma_l$ (drawn on  Fig.\ref{hinh3}) are respectively the rays oriented
from the origin to infinity such that
$$\gamma_{l} = \Big\{q\in \C /\arg(q)=-\frac{\pi}{2} -
\frac{2\pi}{m+2}\Big\}, \hspace{5mm}
\gamma_{r} = \Big\{q\in \C /\arg(q)=-\frac{\pi}{2} +
\frac{2\pi}{m+2}\Big\}.
$$

We now introduce a sesquilinear form on $\mathfrak{H}$ as follows.
For $f,g \in \mathfrak{H}$, we define
\begin{equation}\label{tichvohuong}
    \ll f,g \gg_{\gamma}\, := \int_{\gamma}f(q)\mathcal{PT}g(q)dq =
    \int_{\gamma}f(q)\overline{g(-\overline{q})}dq
\end{equation}
By changing the variable of integration $z = -\overline{q}$, one
easily checks that for every $f,g \in \mathfrak{H}$,
\begin{equation}\label{tichvohuongdx}
    \ll f,g \gg_{\gamma}\, = \overline{\ll g,f \gg}_{\gamma}.
\end{equation}
This means that  $\ll f,g \gg_{\gamma}$ realizes  a
Hermitian sesquilinear form on $\mathfrak{H}$.

We naturally  define the mapping $\|\cdot\|_{\mathcal{PT}}^2 :
\mathfrak{H} \rightarrow \R$ induced by (\ref{tichvohuongdx}):
\begin{equation}\label{chuan}
     \|f\|_{\mathcal{PT}}^2 :=\, \ll f,f \gg_{\gamma}
\end{equation}
Here, it is necessary to emphasize that the Hermitian sesquilinear
form   $\ll ., .\gg_{\gamma}$  has no reason to be positive
definite. In particular,  $\|f\|_{\mathcal{PT}}^2$ can  be zero even
if $f\neq 0$. Hence, it is deservedly  called $\mathcal{PT}$-\emph{pseudo-norm}, while the form  $\ll
., .\gg_{\gamma}$ will be called $\mathcal{PT}$-\emph{inner
product}.

\begin{Pro}\label{}
    The $\mathcal{PT}$-symmetric Hamiltonian $H$ is symmetric
    under the $\mathcal{PT}$-inner product:
    $$\ll H\phi,\psi \gg_{\gamma}\, = \ll \phi,H\psi
    \gg_{\gamma}\, \qquad\qquad \forall\, \phi,\psi \in
    \mathfrak{H}.$$
\end{Pro}
\begin{proof}\label{}
We note that if $\phi$ belongs to $\mathfrak{H}$, then $H\phi \in
\mathfrak{H}$ also.  Now by definition we have
    $$      \ll H\phi,\psi \gg_{\gamma}\,   = \int_{\gamma}H\phi\mathcal{PT}\psi dq
     = \int_{\gamma}(-\phi''+V(q)\phi)\mathcal{PT}\psi dq .$$
    Integrating the right-hand side by parts twice yields
$$\begin{array}{ll}
 \displaystyle
     \ll H\phi,\psi \gg_{\gamma}\,  &
 \displaystyle
= \int_{\gamma}\phi\big( -(\mathcal{PT}\psi)''+ V(q)\mathcal{PT}\psi\big)dq
      \\\vspace{2mm}
       &
\displaystyle
= \int_{\gamma}\phi H\mathcal{PT}\psi dq = \int_{\gamma}\phi \mathcal{PT}H\psi dq = \ll \phi,H\psi \gg_{\gamma}.\\
    \end{array}$$
    This completes the proof.
\end{proof}
 This proposition induces the following two direct consequences.
\begin{Cor}\label{trucgiao}
    Two eigenfunctions corresponding to distinguish \emph{real}
    eigenvalues are orthogonal with respect to $\ll \cdot,\cdot
    \gg_{\gamma}$.
\end{Cor}
\begin{Cor}\label{tinhsuybien}
    If $\phi_E$ is an eigenfunction corresponding to a \emph{complex} eigenvalue $E$ then
    $\|\phi_{E}\|_{\mathcal{PT}}^2 =0$.
\end{Cor}

\subsection*{Stokes multipliers} \label{dapressib} In what
follows we briefly recall some important results of Sibuya
\cite{Sibuya} on complexe second-order linear differential equations. We
consider in the complex plane the following equation
\begin{equation}\label{EquatX}
-\Phi''(X) + W(X)\Phi(X)=0
\end{equation}
where $ W(X) = X^{m} +a_{1}X^{m-1}+\cdots +a_{m}$ is a monic
polynomial function of degree $m \in \N$, with coefficients  $
a_{1},a_{2},\ldots,a_{m} \in \mathbb{C} $.

The following theorem, which is due to Sibuya, asserts the
existence and uniqueness of a solution characterized by its
asymptotic behavior at infinity.

\begin{Thm}\label{ketquaSibuya}
 Equation (\ref{EquatX}) admits a unique solution
 $ \Phi_0(X,a) := \Phi_0(X;a_{1},a_{2},\ldots,a_{m}) $ such that:
\begin{enumerate}
\item $\Phi_0(X,a)$ is an entire function in
                  $(X;a_{1},a_{2},\ldots,a_{m})$,
\item $\Phi_0(X,a)$ and its derivative $ \Phi'_0(X,a) $ admit the
following asymptotic behaviors
\begin{equation}\label{AsymptXprime}
\Phi_0(X) \simeq X^{r_{m}} e^{-S(X,a)} \left[ 1+O(X^{-1/2})\right]
\end{equation}
\begin{equation}\label{tiemcandaoham}
\Phi'_0(X) \simeq X^{\frac{m}{2}+r_{m}} e^{-S(X,a)} \left[
  -1+O(X^{-1/2})\right]
\end{equation}
when $X \rightarrow \infty$ in each sub-sector strictly contained
in the sector
$$ \Sigma_{0} = \left\lbrace \vert \arg(X) \vert <
\frac{3\pi}{m+2} \right\rbrace
$$
and the asymptotic regimes occur uniformly with respect to $a =
(a_{1},a_{2},\ldots,a_{m})$ in any compact of $\C^m$.
\end{enumerate}
\end{Thm}

In the above theorem $r_{m}$ and $S(X,a)$ can be determined
explicitly from $W(X)$. As $X \rightarrow \infty$, one can write
\begin{equation}\label{tam}
\begin{array}{ll}
  \sqrt{W(X)} & = X^\frac{m}{2}\left\{1 + a_{1}X^{-1}+\cdots
  +a_{m}X^{-m}\right\}^{1/2} \\
   & = X^\frac{m}{2}\left\{1+ \sum_{k=1}^{\infty}
 b_{k}(a)X^{-k}\right\}\\
\end{array}
\end{equation}
where, obviously, $b_{k}(a)$ are quasi-homogeneous polynomials in
$a_1,\ldots,a_m$ with real coefficients.

By integrating term-by-term the series in the right-hand side, we
get a primitive of $\sqrt{W(X)}$. The function $S(X,a)$ is
associated to the ``principal part" of this primitive
$$S(X,a) =
\frac{2}{m+2}X^{\frac{m+2}{2}} + \cdots $$
 that only contains
terms with strictly positive powers of $X$. And $r_m = r_m(a)$ is
given by
\begin{equation}\label{dncuarm}
    r_m(a) = \left\{\begin{array}{ll}
      -m/4 & \rm{for}\; $m$\; \rm{odd} \\
       -m/4 - b_{1+m/2}(a) & \rm{for}\; $m$\; \rm{even}\\
    \end{array}\right.
\end{equation}

We should notice that for $m
> 2$, $r_m(a)$ does not depend on the last
coefficient $a_m$ and if all $a_j$ (possibly except $a_m$) are
equal to zero then $r_m =-m/4$.

We shall define other solutions of (\ref{EquatX}) by introducing a
rotation of the complex plan. Let us denote
$$\omega := \exp\left\{-\frac{i2\pi}{m+2}\right\} \, \, \rm{ and }
\, \, \omega_k(a):= (\omega^{k} a_1, \omega^{2k} a_2, \ldots ,
\omega^{km} a_m); \, \, \, (k\in \Z)$$
For each $k\in \Z$, we construct functions $\Phi_k(X;a)$ by
setting
\begin{equation}\label{lesPhik}
\Phi_{k}(X;a) := \Phi_0(\omega^{k}X;\omega_{k}(a)).
\end{equation}

It is not difficult to check that $\Phi_{k}(X;a)$ are indeed
solutions of (\ref{EquatX}) and exponentially vanishing at
infinity in the corresponding sector
\begin{equation}\label{cacsector}
    S_k := \Big\{|\arg(X)-k\frac{2\pi}{m+2}| < \frac{\pi}{m+2}\Big\}.
\end{equation}
The following lemma, which can be verified in a straightforward
way (see \cite{DelabR,Sibuya}), implies the linear independence of
two consecutive solutions $\Phi_k$ and $\Phi_{k+1}$.
\begin{Lem}\label{bode1}
 For any $k \in \mathbb{Z}$, the Wronskian of $\Phi_k$ and $\Phi_{k+1}$ is given by the formula
  \begin{equation}
    \label{Wronskien}
{\sf Wr}(\Phi_{k},\Phi_{k+1}) = 2(-1)^k \omega ^
{-\frac{km}{2}+r_m(\omega_{k+1}(a))}
  \end{equation}
\end{Lem}

From this observation, together with classical results on the
structure of solutions of linear differential equations, we can
infer that $\{\Phi_k,\Phi_{k+1}\}$ constitutes a basis for the
vector space of the solutions of the equation (\ref{EquatX}).
Therefore, every
 solution can be expressed as a linear combination of
$\Phi_k,\Phi_{k+1}$. In particular, for each $k\in \Z$, we have
\begin{equation}\label{bieudien}
    \Phi_{k-1} =C_{k}(a)\Phi_{k}+\widetilde{C}_{k}(a)\Phi_{k+1}
\end{equation}
The multipliers $C_{k}(a)$ and $\widetilde{C}_{k}(a)$ are called the
\emph{Stokes multipliers} of $\Phi_{k-1}$ with respect to
$\Phi_{k}$ and $\Phi_{k+1}$. Further studies on these objects are
addressed in \cite{DelabR,Ph96,Sibuya}. By definition, it is
evident that
$$C_{k}(a)=\frac{{\sf Wr}(\Phi_{k-1},\Phi_{k+1})}{{\sf Wr}(\Phi_{k},\Phi_{k+1})}
\quad {\rm and }\quad \widetilde{C}_{k}(a) =\frac{{\sf
Wr}(\Phi_{k-1},\Phi_{k})}{{\sf Wr}(\Phi_{k+1},\Phi_{k})}
$$
Since $\Phi_k(X;a)$ are entire functions in $a$, it follows
immediately from these equalities and Lemma \ref{bode1} that
$C_k(a)$ and $\widetilde{C}_k(a)$ are also entire functions in
$a$. Furthermore, we also get an explicit expression for
$\widetilde{C}_k(a)$
$$
    {\widetilde{C}}_{k}(a) =  \omega ^
{m+2r_m(\omega_k(a))}
$$
We emphasize that $\widetilde{C}_k(a)$ is never vanishing. Thus, this
coefficient can be reduced to 1 by a suitable renormalisation of the $\Phi_k$'s. For
instance, when $k=0$, (\ref{bieudien}) reads
\begin{equation}\label{ourcase}
    \Phi_{-1} =C_{0}(a)\Phi_{0}+\omega ^
{m+2r_m(a)}\Phi_{1}.
\end{equation}
By setting
\begin{equation}\label{ttt}
\begin{array}{ll}
   & Y_1 := \omega ^ {m/2+r_m(a)}\Phi_{1} = \omega ^
{m/2+r_m(a)}\Phi_0(\omega X;\omega(a)) \\\vspace{2mm}
  {\rm and}\quad & Y_{-1} := \omega ^ {-m/2-r_m(a)}\Phi_{-1} = \omega ^
{-m/2-r_m(a)}\Phi_0(\omega^{-1} X;\omega_{-1}(a))\ , \\
\end{array}
\end{equation}
we can write (\ref{ourcase}) under a slightly symmetric form,
\begin{equation}\label{ourcasedx}
    Y_{-1} =C(a)Y_{0}+Y_{1},
\end{equation}
where $Y_0$ stands for $\Phi_0$ and $C(a) := \omega ^
{-m/2-r_m(a)}C_0(a)$ is also called the \emph{Stokes multiplier} of
$Y_{-1}$ with respect to $Y_0$.

With these conventions, we get a very simple expression for the
Wronskian of $Y_0$ and $Y_1$, namely:
    \begin{equation}\label{wron}
      {\sf Wr}(Y_0,Y_1) = 2.
    \end{equation}
Concerning the (sole) Stokes multiplier $C(a)$ in (\ref{ourcasedx}),
which plays a very important role for our purposes, we have
\begin{Pro}\label{md2} For any $a\in \C^m$,
\begin{equation}\label{hethuc}
             \overline{C(a)} + C(\overline{a}) = 0.
\end{equation}
\end{Pro}

\begin{proof}
By virtue of the quasi-homogeneity of the equation (\ref{EquatX}),
we can see that $\overline{\Phi_0(\overline{X},\overline{a})}$ is
also one of its solutions whose asymptotic behavior
at $\infty$ in the sector $S_0$ is the same  as $\Phi_0(X,a)$.

The uniqueness of the canonical solution in Theorem
\ref{ketquaSibuya} implies immediately that
\begin{equation}\label{doixung}
    \overline{\Phi_0(X,a)} = \Phi_0(\overline{X},\overline{a})
\end{equation}
Taking into account the above definitions of $Y_{-1}$ and $Y_1$,
we can check without difficulty that
\begin{equation}\label{dxcuanghiem}
   Y_{-1}(\overline{X},\overline{a}) = \overline{Y_1(X,a)}
\end{equation}
for any $X\in \C$ and any $a\in \C^m$.

Putting these relations in (\ref{ourcasedx}) leads the desired
identity.
\end{proof}

\subsection*{Eigenvalues as zeros of the Stokes multiplier}

We now consider the complex Sturm-Liouville  eigenvalue problem
associated with the Hamiltonian $H$ in (\ref{hamil}):
\begin{equation}\label{Sturm}
    \left\{\begin{array}{l}
\displaystyle      -\phi''(q)+ V(q)\phi(q)=E\phi(q) \\
\\
      \displaystyle \lim_{q\rightarrow -i\infty.e^{i\theta}}\phi(q) =0
      \quad {\rm and}\quad \lim_{q\rightarrow -i\infty.e^{-i\theta}}\phi(q) =0\\
    \end{array}\right.
\end{equation}
where $\theta := \frac{2\pi}{m+2}$.

 It is necessary to notice that
the boundary condition ($\lim\phi(q)=0$) in (\ref{Sturm}) is
equivalent to the fact that $\phi(q)$ is exponentially vanishing
at infinity in both of the two sectors $S_l$ and $S_r$.

For our purposes, we prefer to consider the problem
(\ref{Sturm}) in a new variable, introducing the rotation  $q \mapsto X :=iq$.
This transforms the differential equation in (\ref{Sturm}) into the
following one\footnote{We can always drop $a_m$  by adding it in $E$.}:
\begin{equation}\label{ptchinh}
    -\Phi''(X) + (X^m + a_1X^{m-1}+\cdots + a_{m-1}X + E)\Phi(X)=0,
\end{equation}
where $\Phi(X)$ stands for $\phi(q)$.

The boundary value conditions in (\ref{Sturm}) turn into the
requirement that the solution $\Phi(X)$ vanishes exponentially in
both of the two sectors $S_{-1}$ and $S_1$, which are defined in
(\ref{cacsector}), see also  Fig.\ref{hinh3}.

With the notations of the previous subsection and writing $C(a,E)
:= C(a_1,\ldots,a_{m-1},E)$ for the Stokes multiplier, we have:

\begin{Lem}\label{khongdiem}  $E_{eigen}$ is an eigenvalue of the problem
(\ref{Sturm}) if and only if $C(a,E_{eigen})
= 0$.
\end{Lem}
\begin{proof}
    Let $\phi_{eigen}(q)$ be an eigenfunction  corresponding to
    the eigenvalue $E_{eigen}$. Then $Y_{eigen}(X) :=
    \phi_{eigen}(-iX)$ solves (\ref{ptchinh}) and vanishes
    exponentially at infinity in both of $S_{-1}$ and $S_1$ in the $X$-plane.
    This fact, together with the observation that $Y_0(X)$ growths exponentially in $S_{\pm
    1}$,
    implies that $Y_{eigen}(X)$ is proportional to
    $Y_{\pm 1}(X)$  in $S_{\pm 1}$ respectively. By virtue of
    (\ref{ourcasedx}),
    this is only possible if $C(a,E_{eigen}) =0$.

    Conversely, if $E_{eigen} = E_{eigen}(a)$ is a zero of $C(a,E)$ then
$Y_{eigen}(X) := Y_{-1}(X,E_{eigen}) \equiv Y_1(X,E_{eigen})$
exponentially vanishes at $\infty$ in both of $S_{\pm 1}$.
Replacing $X = iq$, we obtain a solution for (\ref{Sturm}).
\end{proof}
\begin{Rem}\label{rem1}
    We note  that by construction, $C(a,E)$ is a non-constant entire
    function in both $a$ and $E$. For each fixed $a \in \C^{m-1}$, as an entire
    function of $E$, $C(a,E)$ has the order of
    $\frac{1}{2}+\frac{1}{m}$ (see \cite{Sibuya}).
    Since the order is a non-integral positive number for $m>2$, $C(a,E)$ must have infinitely
    many zeros $E_n = E_n(a)$. By estimating the asymptotic
    behavior of $C(a,E)$ as $E \rightarrow \infty$, Sibuya proved
    that, except possibly for a finite number, these zeros are
    \emph{simple} (i.e. the derivative $\frac{\partial}{\partial
    E}C(a,E) \neq 0$ at those points). Furthermore, the large zeros are known to be close to
    the positive real half-axis and satisfy the following estimate:
    \begin{equation}\label{}
    E_n =
    \Big(\frac{(2n-1)\pi}{2K\sin(2\pi/m)}\Big)^{\frac{2m}{m+2}}\big[1 +
    \nu_n\big]
\end{equation}
where $\displaystyle K := \int_0^{+\infty}(\sqrt{1+t^m}-\sqrt{t^m})dt >0$ and
$\nu_n \rightarrow 0$, $n\rightarrow\infty$.

The most interesting thing here is that the first terms in this
asymptotic estimate do not depend on $a$.

    Under the assumption that all coefficients $a_j$ in (\ref{dathuc})
    are real, the eigenvalues $E_n$ of the
    Hamiltonian $H$ are real or complex conjugate in pairs
    according to Lemma \ref{khongdiem}, as already
    mentioned  in Proposition \ref{doixungxung}. Some
    intensive studies on the reality of all the eigenvalues could be
    found in \cite{Dorey,Shin002}.
\end{Rem}

\section{Indefiniteness of $\mathcal{PT}$-pseudo-norm}

In what follows we shall assume that all the  coefficients $a_j$
are real, so that the  Hamiltonian $H$ defined in (\ref{hamil}) is
$\mathcal{PT}$-symmetric and as a consequence  all the eigenvalues
$E_n$ of the problem (\ref{Sturm}) are real or complex conjugate
in pairs.

 In the sequel, for the sake of simplicity, we do not
always mention the parameter $a$ in the notations, for instance
$C(E)$ will be written in place  of $C(a,E)$.


Let $E_n$ be an eigenvalue of the problem (\ref{Sturm}). Then we
have $C(E_n) =0$ and also $Y_1(X,E_n) \equiv Y_{-1}(X,E_n) =:
Y_{E_n}(X)$. Let $L$ be any endless oriented path in the $X$-plane,
starting from infinity in $S_{-1}$ and then going back to infinity,
but in $S_1$. The following theorem will result from a quite simple
proof.\footnote{We refer the reader to Sibuya's book
(\cite{Sibuya}, ch.6) for a comparison.}

\begin{Thm}\label{dlchinh}
With the above assumptions, we have
$$\displaystyle
\int_LY_{E_n}^2(X)dX = -2C'(E_n)$$ where the prime denotes for the
derivation with respect to the eigenparameter $E$.
\end{Thm}

\begin{proof}\label{}
We start with $Y_1(X,E)$ which  solves the equation
 \begin{equation}\label{cmdlchinh1}
  -Y_1'' + (X^m + a_1X^{m-1}+\cdots +
a_{m-1}X + E)Y_1 = 0.
 \end{equation}
Making $E$ varying, and taking the derivative with respect to $E$, we
immediately deduce that
$Z_1(X,E) := \frac{\partial Y_1}{\partial E}(X,E)$ satisfies
\begin{equation}\label{aaab}
  - (Z_1)_{XX}^{''} + (X^m + a_1X^{m-1}+\cdots +
a_{m-1}X + E)Z_1  + Y_1 = 0
\end{equation}
Combining these equalities together yields
\begin{equation}\label{yyy}
-\big((Z_1)_X^{'}Y_1 - Z_1(Y_1)_X^{'}\big)_X^{'} + Y_1^2 = 0.
\end{equation}
Next, by fixing an arbitrary point $X_0\in L$, we can decompose
$$L = L_1 - L_{-1}\, ,$$
where $L_{\pm 1}$ are oriented path starting from $X_0$ and ending
at infinity in $S_{\pm 1}$ respectively.

Note that both of $Y_1$ and $Z_1$ are exponentially vanishing as
$X \rightarrow \infty$  along $L_1$. Therefore, by integrating
(\ref{yyy}) on the path $L_1$, we obtain
\begin{equation}\label{aaa1}
    \int_{L_1}Y_1^2(X,E)dX = \big((Z_1)_X^{'}Y_1 -
    Z_1(Y_1)_X^{'}\big)\Big|_{X_0}^{\infty\in L_1} =
    {\sf
    Wr}_X(Z_1,Y_1)\Big|_{X=X_0} .
\end{equation}
Similarly, by denoting $Z_{-1}(X,E) := \frac{\partial
Y_{-1}}{\partial E}(X,E)$, we  also have
\begin{equation}\label{aaa2}
    \int_{L_{-1}}Y_{-1}^2(X,E)dX = {\sf
    Wr}_X(Z_{-1},Y_{-1})\Big|_{X=X_0} .
\end{equation}
Substituting $E=E_n$ into (\ref{ourcasedx}) and its derivative
with respect to $E$ yields
$$Y_1(X,E_n) = Y_{-1}(X,E_n)$$  and
$$Z_{-1}(X,E_n) = Z_{1}(X,E_n) + C'(E_n)Y_0(X,E_n) .$$
Now, combining these equalities with (\ref{aaa1})   and
(\ref{aaa2}), one gets
$$
\begin{array}{ll}
 \displaystyle \int_{L}Y_{E_n}^2(X)dX & = \displaystyle\int_{L_1}Y_1^2(X,E_n)dX -\int_{L_{-1}}Y_{-1}^2(X,E_n)dX \\
  \vspace{2mm}
   & =  {\sf
    Wr}_X\big(Z_1(X,E_n) -
    Z_{-1}(X,E_n),Y_{1}(X,E_n)\big)\Big|_{X=X_0}\\\vspace{2mm}
    & = -C'(E_n){\sf Wr}(Y_0(X,E_n),Y_1(X,E_n)).
\end{array}
$$
Taking into account (\ref{wron}), we get the conclusion.
\end{proof}
As a consequence, we now can derive the sign of the
$\mathcal{PT}$-pseudo-norm from the sign  of the derivative of the Stokes
multiplier. Indeed,
 let $E_n$ be a real eigenvalue and $\phi_n(q)$ be an
eigenfunction corresponding to $E_n$. We emphasize that, by the
reality of $E_n$, $\phi_n(q)$ can be chosen to be
$\mathcal{PT}$-symmetric.
\begin{equation}\label{aaa}
\mathcal{PT}\phi_n(q) = \phi_n(q)
\end{equation}
We now have:
\begin{Thm}\label{maintheorem}
        $$\qquad \|\phi_n\|_{\mathcal{PT}}^2 = \ll \phi_n,\phi_n\gg_{\gamma}\; = iK_nC'(E_n),$$
where $K_n$ is a positive real number.
\end{Thm}
\begin{proof}\label{}
Assume that $\gamma$ has been chosen as in (\ref{chongamma}). By
changing the integral variable $iq = X$, we obtain:
$$\|\phi_n\|_{\mathcal{PT}}^2 = \int_{\gamma}\phi_n^2(q)dq =
-i\int_L\phi_n^2(-iX)dX,
$$
where $L$ is the image of $\gamma$ under the mapping $q \mapsto iq
=: X$ (see Fig.\ref{hinh3}).
\begin{figure}[ht]
    \begin{center}
    \rotatebox{0}{\includegraphics[bb=0 0 452 177, scale=.8]{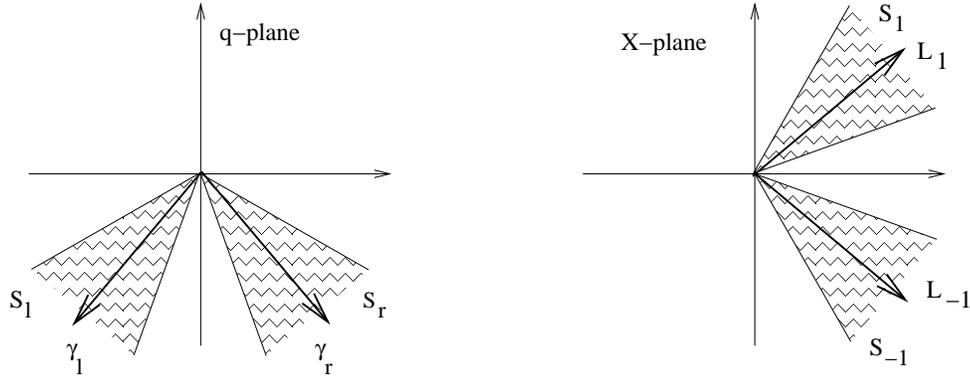}}
    \end{center}
   \vspace{0cm}
    \caption{The $\gamma := \gamma_r - \gamma_l$ and $L := L_1 - L_{-1}$
    in $q$-plane and $X$-plane respectively.}\label{hinh3}
\end{figure}

Since $\phi_n(q)$ is an eigenfunction, $\phi_n(-iX)$ must
 vanish exponentially at infinity in both $S_{\pm 1}$. So, there
 exists a non-zero  constant $\alpha_n$
such that
$$\phi_n(-iX)=\alpha_n Y_1(X,E_n) = \alpha_n Y_{-1}(X,E_n)$$
When $X$ take real values, we can deduce from (\ref{dxcuanghiem})
 and (\ref{aaa}) that $Y_{\pm 1}(X,E_n)$ and $\phi_n(-iX)$ are all real.
 Hence, $\alpha_n \in \R$.

By Theorem \ref{dlchinh}, we have
$$\|\phi_n\|_{\mathcal{PT}}^2 = i2\alpha_n^2C'(E_n)$$
The proof is complete by setting $K_n := 2\alpha_n^2$.
\end{proof}
The following result is a direct consequence of the theorem:
\begin{Cor}\label{hequa3}
Assume that $E_n$ is a real eigenvalue of $H$, then
    $\|\phi_n\|_{\mathcal{PT}}^2 = 0$ if the multiplicity of $E_n$
    is greater than $1$.
\end{Cor}
We now remind a classical result from real analysis:
\begin{Lem}\label{giaitichthuc}
    Suppose that a real-valued function $f(x)$ is continuously
    differentiable on $(a,b)$ and $x_1, x_2 \in (a,b)$ are  its two consecutive
    zeros. Then $f'(x_1)f'(x_2) \leq 0$.
\end{Lem}
\begin{proof}\label{}
    Assume conversely that $f'(x_1)f'(x_2) > 0$; for instance
    $f'(x_1) >0$ and $f'(x_2)>0$. We deduce from this assumption that
    $f(x)$ increases  strictly in sufficiently small neighborhoods
    of $x_1$ and $x_2$. This implies that $f(x_1 +\epsilon) >0$
    and $f(x_2 -\epsilon) <0$ for a sufficiently small $\epsilon
    >0$. By its continuity, $f(x)$ must vanish at least once
     in the interval $(x_1+\epsilon,x_2 -
     \epsilon)$. This is contrary to the hypothesis.
\end{proof}
This lemma asserts that if all zeros $x_n$ of $f(x)$ are
simple then $f'(x_n)$ changes its sign alternatively. Hence a switch
like $(-1)^n$. This is exactly what we describe in the  next
theorem.
\begin{Thm}\label{dinhlyphu}
    Assume that all eigenvalues $E_n$ of the problem (\ref{Sturm}) are
    \emph{real} and \emph{simple}. Then, up to a normalization, the
    set of the corresponding eigenfunctions $\{\phi_n\}, n\geq 0$,
 is $\mathcal{PT}$-orthonormal in
    the sense that
    \begin{equation}\label{dinhlyphu1}
      \ll \phi_n , \phi_m\gg_{\mathcal{PT}}\; = \pm (-1)^n\delta_{mn}
    \end{equation}
where $\pm 1 = \ll \phi_0 , \phi_0\gg_{\mathcal{PT}}$.
\end{Thm}
\begin{proof}\label{}
    We first notice that the reality of the whole set of eigenvalues $\{E_n\}$
    enables us to treat $C(E) := C(a,E)$ as a function of the real variable $E$
    (for each fixed $a \in \R^{m-1}$). This function is purely
    imaginary because of (\ref{hethuc}). Hence, $-iC(E)={\rm Im}C(E)$ is a real-valued
    function of $E \in\R$. So also is $-iC'(E)$.

    For $m\neq n$, the equality (\ref{dinhlyphu1}) is a direct
    consequence of Corollary    \ref{trucgiao}. We should note that, as $m = n$,
the sign of $\|\phi_n\|_{\mathcal{PT}}^2$ does not change when
multiplying $\phi_n$ by a non-zero constant. Therefore, by
applying  lemma
    \ref{giaitichthuc} to $-iC(E)$ and normalizing $\phi_n$, we get
    the conclusion.
\end{proof}

We emphasize that the hypothesis on the simpleness of all
eigenvalues in this theorem is crucially necessary. Nevertheless,
this requirement is not always satisfied, especially in cases of
Hamiltonians involving parameters.

To end this section, we now  detail the striking
case when all $a_j = 0$ and $m \geq 2$. The Hamiltonian then
becomes $H = p^2 + q^2(iq)^{m-2}$, whose whole eigenvalues have been
proved to be  real and positive using several different
methods (see \cite{BenBoe98,BenBoe99,DP98-1,DT00,Dorey,Shin002}).

In their recent papers \cite{Ben0,Ben1}, Bender \emph{et al}
showed that for some unbroken $\mathcal{PT}$-symmetric
Hamiltonians, the charge operator $\mathcal{C}$ can be computed
explicitly using perturbative techniques. They also provided
numerical evidences on the sign alternation of
$\|\phi_n\|_{\mathcal{PT}}^2$. We can now  explain this phenomenon
quite simply.

Let's agree on the simpleness of all eigenvalues $E_n$, which will be
the matter of a forthcoming paper, so that each $-iC'(E_n)$ possesses
a sign. Note that by construction, the entire function $Y_0(X,E)$
is the solution (unique by its asymptotic behavior at infinity)
of equation (\ref{ptchinh}) which vanishes exponentially at
$+\infty$ and takes only real-values whenever $X$ and $E$ are
real. It is not hard to verify that $Y_0(0,0) \neq 0$. Therefore,
substituting $X=0$ and $E=0$ in (\ref{dxcuanghiem}) yields
$$\omega ^
{-m/4}Y_0(0,0) = C(0)Y_0(0,0) + \omega ^ {m/4}Y_0(0,0)
$$
This implies immediately that
$$-iC(0) = \frac{\omega^{-m/4} - \omega ^{m/4}}{i} = 2\sin\Big(\frac{2m\pi}{2(m+2)}\Big) > 0$$
One can regard $-iC(E)$ now as a (real-valued) function of $E \in
\R$ with only simple zeros $0 < E_0 < E_1 < \cdots $ and starting
with a positive value $-iC(0) > 0$. As a direct consequence of
lemma \ref{giaitichthuc}, we obtain
$${\rm sign}(-iC'(E_n)) =  (-1)^{n+1}\ ; \quad \forall n \in \N $$
Then after normalizing $\phi_n$, we  finally find out
$$\|\phi_n\|_{\mathcal{PT}}^2 = (-1)^{n}\, ,\qquad {\rm for } \quad n = 0,1,2,3,\ldots$$

\section{Conclusion}

In this paper, we have established an explicit relation between
$\mathcal{PT}$-pseudo-norm and the derivative of the Stokes multiplier $C(E)$
with respect to the eigenparameter. In this formulation, the
indefiniteness of the $\mathcal{PT}$-pseudo-norm associated with a
non-Hermitian but $\mathcal{PT}$-symmetric  Hamiltonian   is
interpreted as a natural sign alternation  of the derivative of
an entire function at its zeros. This relation also supplies us a
simple criterion to recognize degenerate eigenstates in the sense
that their $\mathcal{PT}$-pseudo-norms are vanishing.

By Corollary \ref{hequa3}, one has to acknowledge the unavoidable
occurrence of such degenerate eigenstates, even if the
corresponding eigenvalues are real. The reality of the whole spectral set
of a $\mathcal{PT}$-symmetric Hamiltonian does not completely
ensure  the non-degeneracy of its eigenstates. This observation
indicates that the unbroken $\mathcal{PT}$-symmetry is not exactly
equivalent to the reality of the whole set of eigenvalues of a
$\mathcal{PT}$-symmetric Hamiltonian.

To strengthen our arguments on this point, we should refer to an
earlier joint work with Delabaere \cite{DT00}, where the
spectrum of the Hamiltonian $H_{\alpha} = p^2+i(q^3 + \alpha q)$ is
investigated by means of (exact) semiclassical analysis. This Hamiltonian exhibits
a version of a degeneracy of its eigenvalues $E_n = E_n(\alpha)$ for
negative real $\alpha$. More precisely, when $\alpha$ goes to
$-\infty$, some pairs of real eigenvalues gradually turn into
complex conjugate. For the first critical value of $\alpha$ where this
phenomenon occurs, the  corresponding pairs of eigenvalues are
nothing but double zeros of $C(E)$, which  means that $C'(E)=0$ at these
zeros. Such a critical value has been computed in \cite{Handy1}
$\alpha_{crit} \simeq -2.6118094$, and the eigenvalues becoming
complex conjugate are the two lowest ones (see \cite{DT00}, Fig.1). We
should notice that in this situation, the eigenvalues of
$H_{\alpha_{crit}}$ are still all real.

Obviously, the action of the charge operator $\mathcal{C}$ (if any
exists) on eigenfunctions in case of degeneracy may be not merely
to switch signs by multiplying $\phi_n$ by $(-1)^n$ as in
\cite{Ben1,BenBerry003}. Nevertheless, since there are at most a
finite number of such degenerate eigenfunctions, we believe that
an analogous method as those in the above citations could be still
applicable to define the charge operator $\mathcal{C}$.

Findings on the indefiniteness of $\mathcal{PT}$-pseudo-norm enable us
to keep on constructing the mathematical apparatus for
$\mathcal{PT}$-symmetric quantum mechanics. In this study, the
structure of a Krein space may be involved.

Moreover, there seems to exist some hidden relations between the
Stokes multiplier $C(E)$ and the calculation of the charge
operator $\mathcal{C}$. At least as we indicated, the simpleness
and the reality of all the zeros of $C(E)$ first lead to the
non-degeneracy of $\mathcal{PT}$-pseudo-norm, which is a necessary
condition to construct this operator.

\vspace*{5mm}

\subsection*{Acknowledgments.}
This work was supported by the Abdus Salam International Centre
for Theoretical Physics (ICTP) in the framework of Post-doctoral
Fellowship.



\begin{thebibliography}{plain}
\begin{footnotesize}


\bibitem{Bag01} B. Bagchi, C. Quesne, M. Znojil, {\it
Generalized Continuity Equation and Modified Normalization in
PT-Symmetric Quantum Mechanics.\/} {\sl
Mod. Phys. Lett. A16 (2001) 2047-2057.}

\bibitem{Ben0}   C.M.~Bender, D.C.~Brody; H.F.~Jones,
   {\it  Must a Hamiltonian be Hermitian?
 Amer. J. Phys. 71 (2003), no. 11, 1095--1102.}

\bibitem{Ben1}
C.M.~Bender, J.~Brod, A.~Refig, M. E Reuter, {\it The  operator
$\mathcal{C}$ in $\mathcal{PT}$-symmetric quantum theories  \/}
{\sl J. Phys. A: Math. Gen. 37 No 43 (2004) 10139-10165.}

\bibitem{BenBerry002}   C.M.~Bender, M. Berry, A.  Mandilara,
{\it Generalized $\mathcal{PT}$ symmetry and real spectra.  \/}
{\sl J. Phys. A 35 (2002), no. 31, L467--L471.}

\bibitem{BenBerry003}  C.M.~Bender, P.N.~Meisinger, Q.Wang, {\it Calculation
of the hidden symmetry operator in
$\mathcal{PT}$-symmetric quantum mechanics. \/} {\sl J. Phys. A 36
(2003), no. 7, 1973--1983.}

\bibitem{BenBoe98}   C.M.~Bender, S.~Boettcher, {\it Real spectra
in non-Hermitian hamiltonians having {\cal PT}-symmetry. \/} {\sl
Phys.~Rev.~Lett. 80, 5243 (1998).}

\bibitem{BenBoe99}   C.M.~Bender, S.~Boettcher, P.N.~Meisinger,
{\it {\cal PT}-symmetric quantum mechanics. \/} {\sl
J.~Math.~Phys. 40, 2201 (1999).}


\bibitem{BenBoe98-2} C.M.~Bender, K.A.~Milton,
{\it  Model of supersymmetric quantum field theory with broken parity symmetry.
 \/} {\sl Physical Review  D,
Vol 57,  N$^o$ 6,   3595-3608 (1998)}

\bibitem{BW69} C.M.~Bender, T.T.~Wu,  {\it Anharmonic oscillator.\/}
{\sl Phys.~Rev. 184, 1231-1260 (1969).}

\bibitem{Boas} R. Ph. Jr. Boas,
{\it Entire functions.  \/}{\sl
Academic Press Inc., New York, 1954.}

\bibitem{CGM80}
E.~Caliceti, S.~Graffi, M.~Maioli,
{\it Perturbation theory of odd anharmonic oscillators.\/} {\sl
Commun.~Math.~Phys. 75, 51-66 (1980).}


\bibitem{CanJunkTrost98} F.~Cannata, G.~Junker, J.~Trost, {\it
Schr\"odinger operators with complex potential but real spectrum.\/}
{Phys. Lett. A 246, 219-226 (1998).}



\bibitem{DP97} E.~Delabaere, F.~Pham,
{\it Unfolding the quartic oscillator.}
{\sl Annals of Physics 261, N$^o$. 2, 180-218 (1997).}

\bibitem{DDP97} E.~Delabaere, H.~Dillinger, F.~Pham, {\it Exact
semi-classical expansions  for one dimensional quantum oscillators.\/} {\sl
Journal Math. Phys. 38, 12, 6126-6184 (1997).}


\bibitem{DP98-1} E.~Delabaere, F.~Pham,
{\it Eigenvalues of complex hamiltonians with $\mathcal{PT}$-symmetry
  I.\/} {\sl Phys.~Lett.~A 250, 25 (1998).}

 \bibitem{DP98-2} E.~Delabaere, F.~Pham,
{\it Eigenvalues of complex hamiltonians with $\mathcal{PT}$-symmetry
  II.\/} {\sl Phys.~Lett.~A 250, 29 (1998).}

\bibitem{DT00} E.~Delabaere, D.T.~Trinh,
{\it Spectral analysis of the complex
cubic oscillator. \/} {\sl J.Phys. A: Math. Gen. 33 (2000),  8771-8796.}

\bibitem{DelabR} E.~Delabaere, J.-M.~Rasoamanana,
{\it Resurgent deformations for an ordinary differential equation
of order 2. \/} {\sl To appear in Pacific Journal of Mathematics.}

\bibitem{Dorey} P. Dorey, C. Dunning, R. Tateo,
{\it Spectral equivalences, Bethe ansatz equations, and reality
  properties in ${\cal PT}$-symmetric quantum mechanics.\/}
{\sl  J. Phys. A 34 (2001), no. 28, 5679--5704. }


\bibitem{Fer}
F.M.~Fern\'andez, R.~Guardiola, J.~Ros, M.~Znojil,
 {\it   A family of complex potentials with real spectrum. \/} {\sl
J. Phys. A 32, N$^o$. 17, 3105-3116  (1999).}


\bibitem{Handy1} C.R. Handy, {\it
Generating converging bounds to the (complex) discrete states of the $P\sp 2+iX\sp 3+i\alpha X$ Hamiltonian.
 \/}{\sl J. Phys. A 34 (2001), no.
24, 5065--5081 }



\bibitem{Japaridze} G.S. Japaridze,  {\it
Space of state vectors in  ${\cal PT}$ symmetrical quantum mechanics
 \/} {\sl  J. Phys. A 35 (2002), no. 7, 1709--1718.}



\bibitem{LevaiZnojil} G. L\'evai, M. Znojil, {\it
 Conditions for complex spectra in a class of
${\cal PT}$-symmetric potentials. \/}{\sl  Modern Phys. Lett. A 16
(2001), no. 30, 1973--1981. }

\bibitem{LevaiCan} G. L\'evai, F.Cannata, A.Ventura, {\it $\mathcal{PT}$-symmetry breaking and explicit
expressions
      for the pseudo-norm in the Scarf II potential. \/}{\sl Phys.Lett.A
300 (2002) 271--281.}


\bibitem{Mez00}   G.A~Mezincescu,
{\it Some properties of eigenvalues and eigenfunctions of
the cubic oscillator
with imaginary coupling constant. \/} {\sl
J. Phys. A. 33 (2000).}

\bibitem{Mostaf}  A. Mostafazadeh,
{\it Pseudo-Hermiticity versus $PT$ symmetry: the necessary condition for
the reality of the spectrum of a non-Hermitian
Hamiltonian. \/} {\sl
J. Math. Phys. 43 (2002), no. 1, 205--214.}

\bibitem{Mostaf2} A. Mostafazadeh, {\it Exact $\mathcal{PT}$-symmetry is equivalent to Hermiticity
.\/} {\sl   J. Phys. A: Math. Gen. 36 No 25 (2003) 7081-7091.}

\bibitem{Mostaf3} A. Mostafazadeh, A.Batal, {\it Physical aspects
of pseudo-Hermitian and $\mathcal{PT}$-symmetric quantum mechanics
.\/} {\sl   J. Phys. A: Math. Gen. 37 No 48 (2004) 11645-11679.}


\bibitem{Ph96}  F.~Pham,
{\it  Confluence of turning points in exact WKB analysis.\/} {\sl
 The Stokes phenomenon and Hilbert's
16th problem (Groningen, 1995), 215--235, World Sci. Publishing, River Edge,
NJ (1996).}

\bibitem{Ph00}  F.~Pham,
{\it  Multiple turning points in exact WKB analysis
(variations on a theme of Stokes).\/}
 {\sl Toward the exact WKB analysis of differential equations linear
   or non-linear (C.Howls, T.Kawai, Y.Takei ed.), Kyoto University
   Press (2000),71-85.}

\bibitem{Shin002} K.C. Shin, {\it On the reality of the eigenvalues for a
    class of  ${\cal PT}$-symmetric oscillators.\/}
{\sl   Comm. Math. Phys. 229 (2002), no. 3, 543--564.}


\bibitem{Sibuya}
 Y. Sibuya,
 {\it Global Theory of a Second Order Linear Differential Equation with a Polynomial Coefficient.\/}
 {\sl Mathematics Studies 18,
 North-Holland Publishing Company (1975).}



\bibitem{These} T.D.~Tai, {\it Asymptotique et
analyse spectrale de l'oscillateur cubique. \/}
{\sl Th\`ese, Universit\'e de Nice (2002).}

\bibitem{Tai2} T.D.~Tai, {\it On the Sturm-Liouville problem for the complex cubic oscillator. \/}
{\sl Asymptotic Analysis, 40 (2004), no. 3-4, 211-234.}

\bibitem{Vo83} A.~Voros, {\it The return of the quartic
    oscillator. The complex WKB method\/} {\sl
Ann. Inst. H.Poincar\'e, Physique Th\'eorique, 39, 211-338 (1983).}

\bibitem{Vor99-2} A. Voros, {\it Exact resolution method for general $1D$
polynomial Schr\"odinger equation.\/} {\sl J. Phys. A 32, N$^o$. 32,
5993--6007 (1999).}


\bibitem{Znoj0104}   M.~Znojil,
{\it  Conservation of pseudo-norm in {\cal PT} symmetric quantum
mechanics. \/} {\sl  Rend. Circ. Mat. Palermo (2) Suppl. No. 72
(2004), 211--218.}

\bibitem{Znoj01}   M.~Znojil,
{\it ${\cal PT}$-symmetrized supersymmetric  quantum mechanics. \/}
{\sl
DI-CRM Workshop on Mathematical Physics (Prague, 2000). Czechoslovak
  J. Phys. 51 (2001), no. 4, 420--428.}

\end{footnotesize}

\end{thebibliography}
\end{document}